# Comment on "Assessing CN earthquake predictions in Italy" by M. Taroni, W. Marzocchi, P. Roselli

G. Molchan[1], A. Peresan[2,3], G.F. Panza[3,4,5], L. Romashkova[1], V. Kossobokov[1,3]

[1] Institute of Earthquake Prediction Theory and Mathematical Geophysics, Russian Academy of Sciences, Moscow, Russian Federation
[2] National Institute of Oceanography and Experimental Geophysics. CRS-OGS, Udine, Italy
[3] International Seismic Safety Organization, Arsita, Italy
[4] Institute of Geophysics, China Earthquake Administration, Beijing, People's Republic of China
[5] Accademia Nazionale dei Lincei, Roma. Italy

The paper by Taroni et al. (2016) considers results of forward prediction of Italian strong earthquakes, during the period 1998-2016, based on CN algorithm. The declared intent of the paper is to give "a *careful* assessment of CN prediction performances … using *standard* testing procedures." This is unlikely feasible goal, however, because the target earthquake data related to each individual CN sub-region of Italy (see Table 3 in the paper) are very limited. Namely, the number N of target events within each region is:

$$N = 5 \,(M>5.3, \text{North}) ;\ 3 \,(M>5.5, \text{Center});\ \text{and}\ 1 \,(M>5.5, \text{South})$$

This situation is non statistical, and a priori it is clear that the *standard* statistical methods are not effective here.

Let us consider the best case, from statistical point of view, provided by North region with 5 target events. Here CN has qualitatively good result: 4 successes out of 5 target events. Formally such result corresponds to the observed significance level *alpha*=5.8% (p-value=0.058) in random guessing, with the success probability 0.357 (Table 3 in the paper). Based on these values the authors conclude that "the model CN and the Poisson model have the same predictive performances". This conclusion needs comments:

1) The *alpha* estimate is unstable over the time, because the number of target events is small. In fact, the next target earthquake in the region will change the score 4/5 as follows: either 5/6, or 4/6. As a result the estimate *alpha* will become 2.4% or 12.5% respectively. Accordingly, in the first case the authors will come to the directly opposite conclusion. This instability is the consequence of the authors' choice to analyze sub-regions with very few data.

2) The *alpha* is nothing more than observed *significance* of the result 4/5 given percentage of space-time in alarm 35.7% (Table 3), and it doesn't represent the *prediction ability* of CN. In this connection, it is useful to consider the standard *prediction ability index*, i.e. a fraction of non-randomly predicted events

$$e = \dot{n} \,(\text{hit rate}) - \tau \,(\text{alarm rate}) \qquad (1)$$

Setting aside the problem of small number of target events, we get

$$e = 4/5 - 0.357 = 44\%$$

It is worthy of note that such estimation of *e*, as large as 44%, is an extremely high value when dealing with the prediction of strong earthquakes. For example, $e \approx 20\%$ for M8 algorithm in the prediction magnitude 8 or larger events worldwide (Molchan and Romashkova, 2010).

3) The values of *alpha* and *e* are simply point estimations. Therefore, to judge the predictive performance of the CN method the interval estimations are necessary, especially in the case of deficiency of data.

4) As an alternative to the above mentioned classical approach, the authors consider also a gambling approach, suggested by Zhuang (2010) and applied recently by Zechar and Zhuang (2014). Their Pari-mutuel Gambling score (PGS) method applied to earthquake forecasting has been analysed in detail by Molchan (2016). Taroni et al. (2016) adapt the PGS approach to the analysis of the alarm-based CN prediction algorithm. The conclusion about predictive ability of the CN method in this case is based on the summary Pari-mutuel Gambling score $W_T$ (computed according to formula (2) of their paper). To explain this quantity, we have to introduce some notations.

Let's represent the period of CN monitoring $T$ as a union of subintervals $\Delta_i$ of length $\Delta$. We define $v_i = +(-)$, when target event happens (not happens) during small time interval $\Delta_i$. Similarly, $A_i = +(-)$ when alarm happens (not happens) in $\Delta_i$. The result of prediction of target events during the period $T$ in total is given by the confusion matrix

$$\begin{pmatrix} n_{++} & n_{+-} \\ n_{-+} & n_{--} \end{pmatrix}$$

where $n_{\alpha\beta} = \#\{\Delta_i : v_i = \alpha, A_i = \beta\}, \alpha, \beta = +(-)$.

In particular, $n_{++} + n_{+-} = N_e$ is total number of target evens, $(n_{++} + n_{-+})\Delta = T_A$ is total alarm time, and $T_A / T = \tau$ is the observed alarm rate.

If $p_\Delta$ is a probability of occurring of target event during $\Delta_i$ in random model, then

$$W_T = [n_{++}(1-p_\Delta)/(1+p_\Delta) - n_{+-}] + [n_{--}p_\Delta/(2-p_\Delta) - n_{-+}] := W_T^+ + W_T^-, \quad (2)$$

This value is interpreted as the gain of a forecaster against random guessing. Therefore, negative values of $W_T$ for a forecaster vote in favor of random guessing. The larger absolute value $W_T$ the stronger the advantage of random guessing.

Usually, to characterize the prediction ability of some method, two statistics are used: the hit rate, $\dot{n} \approx n_{++} / N_e$, and the rate of alarm, $\tau$ (Molchan, 1997).

In the Pari-mutuel Gambling method the concept of success is interpreted more broadly: success is counted as a correct prediction of target event or his absence (quiescence) in a given subinterval $\Delta_i$. The successes in prediction of events, $n_{++}$, and of quiescence, $n_{--}$, are equally presented in the statistics. In fact, $W_T^+$ and $W_T^-$ are the components of the total gain of a forecaster due to successes in prediction of event and of no event, correspondingly. Formulas for $W_T^+$ and $W_T^-$ are identical and differ only by the probabilities of the predicted phenomenon in random model: Pr(event)= $p_\Delta$ and Pr(no event)= $1-p_\Delta$.

Equal accounting of objects of different types: target events (points) and no target events (intervals $\Delta_i$), is a crucial point for understanding of properties of the statistics $W_T$.

By (2), the gain $W_T^+$, associated with the "art of prediction" of target events, is limited for any bin size $\Delta$: $W_T^+ < n_{++} - n_{+-}$ ($\leq 3$ for any of the three sub regions). Note, that any smooth score, based on ($n_{++}, n_{+-}, \tau$), also is stable as $\Delta \to 0$, because any target event is a point object.

The situation with $W_T^-$ is different. In this case, the numbers of successes $n_{--}$ and failures $n_{-+}$ are increasing with $\Delta \to 0$. However, in the random model the case of no target event occurrence in a small interval $\Delta$ is highly probable and therefore dividends from its prediction, according to (2), are limited:
$$n_{--} p_\Delta / (2 - p_\Delta) < n_{--} \Delta \lambda < T \lambda \quad (\leq 5 \text{ in our situation}),$$
where $\lambda$ is the rate of target events.

At the same time, the penalty for prediction error unlimitedly grows
$$n_{-+} \approx T_A / \Delta, \quad \Delta \to 0.$$
As a result, the total gain in prediction of target event (i.e. successful prediction) and no target event (i.e. no alarm and no earthquake occurrence) is determined largely by the value of $-T_A / \Delta$. To test this theoretical conclusion, let us put $T = 224$ months (Table 1), and $\Delta = 2$ months (time bin in CN algorithm application). Then $-W_T \approx 112\tau$ regardless of the hit rate. Table 1 (compiled based on Tables 3 and 4 by Taroni et al. (2016)) fully confirm this conclusion:

| Zone | North | Center | South |
|---|---|---|---|
| $\tau$, in % | 35.7 | 35.7 | 25.0 |
| $W_T$ | -32.7 | -36.7 | -27.7 |
| Hit rate | 4/5 | 2/3 | 0/1 |

Table 1 - Values of Percentage of space-time in alarm $\tau$, Parimutuel Gambling score $W_T$ and Hit rate, as reported in Tables 3 and 4 of the paper by Taroni et al. (2016)

Note that the generation of the space-time alarms is an intelligent essence of the prediction algorithms like CN or M8. And as we have shown, this essence is penalized at the highest degree by the PGS approach, because almost each bin of the alarm is interpreted as an error.
Therefore, one can conclude that the results of the analysis of CN algorithm on the base of the Pari-mutuel Gambling score are irrelevant to assessing its prediction performance. The estimations of $W_T^\pm$ show that, under the condition
$$2\lambda\Delta << \tau, \tag{3}$$
the statistic $W_T$ will provide a negative verdict about significance of any time prediction algorithm with arbitrary number of target events. To be clear, (3) means that the average time interval between target events is much larger than the time step of updating the alarm $\Delta$, which is the case for CN algorithm.

*Conclusions*

A very limited amount of data is a serious obstacle for statistical analysis of CN prediction algorithm at the regional level of Italy. The attempt to replace the standard approaches by Pari-mutuel Gambling method leads to almost complete loss of information about predicted earthquakes, even for a large sample of target events. Therefore, the conclusions based on PGS, are untenable. As noted by Zhuang (2016, personal communication) "It seems to me that forecasting and betting should be separated". An in-depth discussion is provided in Molchan (2016) and, much earlier, in Molchan and Romashkova (2011).